# Ultrafast phononic switching of magnetization

A. Stupakiewicz[1*], C. S. Davies[2,3], K. Szerenos[3], D. Afanasiev[4], K. S. Rabinovich[5], A. V. Boris[5], A. Caviglia[4], A. V. Kimel[3], and A. Kirilyuk[2,3*].

[1]*Faculty of Physics, University of Bialystok, 1L Ciolkowskiego, 15-245 Bialystok, Poland.*
[2]*FELIX Laboratory, Radboud University, 7c Toernooiveld, 6525 ED Nijmegen, The Netherlands.*
[3]*Radboud University, Inst. for Molecules and Materials, 135 Heyendaalseweg, 6525 AJ Nijmegen, The Netherlands.*
[4]*Kavli Institute of Nanoscience, Delft University of Technology, Lorentzweg 1, 2628 CJ Delft, The Netherlands.*
[5]*Max Planck Institute for Solid State Research, Heisenbergstraße 1, 70569 Stuttgart, Germany.*

*Correspondence to: and@uwb.edu.pl (A.S.); andrei.kirilyuk@ru.nl (A.K.).

**Identifying an efficient pathway to change the order parameter via a subtle excitation of the coupled high-frequency mode is the ultimate goal of the field of ultrafast phase transitions[1,2]. This is an especially interesting research direction in magnetism, where the coupling between spin and lattice excitations is required for magnetization reversal[3,4]. Despite several attempts[5,6] however, the switching between magnetic states via resonant pumping of phonon modes has not yet been demonstrated. Here we show how an ultrafast resonant excitation of the longitudinal optical phonon modes in magnetic garnet films switches magnetization into a peculiar quadrupolar magnetic domain pattern, unambiguously revealing the magneto-elastic mechanism of the switching. In contrast, the excitation of strongly absorbing transverse phonon modes results in thermal demagnetization effect only.**

In condensed matter systems, the collective excitation modes (phonons, magnons, phasons, spin fluctuations, etc…) define the energy range that determines all the important and intriguing thermodynamic and macroscopic properties of solids, such as electric, magnetic, or crystallographic order, superconducting transition temperatures and so on. Sufficiently strong resonant excitation of these modes, at the appropriate energy, can efficiently induce a profound modification of macroscopic properties, including a permanent change in the order parameter[1-6].

Control of the crystal environment arguably represents the most universal mechanism to act on magnetization, as it is present in all materials regardless of their magnetic structure. The emerging field of "straintronics" aims to study how mechanical strains of the atomic lattice in a solid respond to extrinsic forces[7-9]. One interesting approach to generate such strains is via the anharmonic interaction of different phonon modes[10], that transfers an excitation of an infrared-active mode into a coordinate shift along a coupled Raman-active coordinate. This shift represents, for the length of the excitation (or for the duration of the phonon lifetime), a change of the crystal structure and as such, a new state of the material. Such "nonlinear phononics"[11] promises a new direction for the ultrafast optical control of solids in their electronic ground state, including for example ultrafast steering of the metal-insulator phase transitions[12], or (temporarily) changing superconducting[13] or ferroelectric[14,15] properties. The most important milestone on the roadmap of nonlinear phononics, that has been inferred to exist but has remained elusive until now, is an ultrafast strain-induced switching of magnetization.

In our study we used yttrium iron garnet (YIG) thin films on a GGG(001) substrate (see materials and methods). In this cubic crystal, strong magnetocrystalline anisotropy renders the cube diagonals the easy directions of magnetization. There are thus eight metastable states for the magnetization vector, as numbered in Fig. 1a. To observe the magnetic domain structures in the sample, we used a magneto-optical Faraday microscope at room temperature (see Methods and Extended Data Fig. S1). Details about the methods of detection and identification of different magnetic domains in these garnet films have been discussed before[16-18].

To provide resonant excitation of the phonon modes in YIG, we sourced pulses from FELIX (Free Electron Lasers for Infrared eXperiments, Nijmegen, The Netherlands). The IR beam with photon energy ranging between 41 meV and 124 meV (wavelength 10–30 µm) was focused to a spot of diameter ~200 µm on the surface of the YIG film. The pulses of FELIX have been shown to be Fourier-transform limited[19], with their bandwidth experimentally tunable in the range of 0.5−2.0%. Various pulse trains of different repetition frequencies were used, from 10 µs-long 'macropulses' at 25 MHz repetition rate, to 10 ns-long bunches of about 10 pulses each, and finally to single sub-picosecond 'micropulses'. The results obtained by these grossly different excitations were qualitatively identical.

Before irradiation, the sample magnetization was saturated along the diagonal [1-11] axis, and two type of magnetic phases (8 and 4 states in Fig. 1a) were observed at zero magnetic field with opposite direction of polar magnetization components and the same in-plane magnetization components (see Fig. 1b). After irradiating the garnet film with light of wavelength 14 µm, four large magnetic domains are formed with their magnetization directions along the other diagonals of the structure (see Fig. 1c, d). We note that for this particular image, a 10 ns long sequence of FELIX micropulses was used. This allowed us to keep the optical peak intensity safely below the damage threshold while simultaneously obtaining domains large enough to be stable. Similar effect of the accumulation of the effect of multiple pump pulses was used in Ref. 18. We have also repeated the experiments using single sub-picosecond micropulses of FELIX, and very similar switching patterns were obtained, albeit stronger focusing and peak intensities approaching damage threshold was required. The validity of the multipulse approach is supported by the experiments showing that the only effect of repeating single micropulses (on the time scale of seconds) is a slight growth of the domains, if any. The micromagnetic simulations (see below) fully support this conclusion, showing only low-amplitude dynamics appearing after pulse repetition, and no difference in the final pattern. To erase a domain pattern with the switched magnetization and to initialize the initial magnetic state, an external in-plane magnetic field of about 10 mT was applied for a short time.

Aiming to reveal the mechanism responsible for the magnetic switching, we tuned the photon energy to different spectral lines while keeping the same pump fluence. The normalized switched area (see Extended Data Fig. S2) obtained from the images of large magnetic domains is plotted as a function of the pump wavelength within the range 10−30 µm (see Fig. 2a). The complex structure of the garnet lattice with large number of ions per Bravais unit cell allows an extended set of infrared active phonon modes in this spectral range[20,21]. We have used synchrotron-based infrared ellipsometry to accurately determine the phonon spectra of the garnet film (see Methods and Extended Data Fig. S3). This method is very sensitive to thin-film properties because of the oblique incidence of light[22], and it yields both the absorption spectrum of infrared-active transverse optical (TO) phonons and the loss function spectrum. The spectral dependence of longitudinal phonon modes (LO) as derived from the ~~optical~~ loss function reveals a pronounced resonant behavior around wavelengths 14 µm and 22 µm (730 cm$^{-1}$ and 445 cm$^{-1}$). It is clear that the behavior of the switched area is extremely well correlated with these LO resonances. The somewhat smaller effect in the vicinity of 23 µm can be easily explained by the ~25% absorption of infrared (IR) light by water vapour during our experiments performed in air. In addition, we have verified that the polarization of the pump does not influence the observed switching, which additionally proves the longitudinal character of the excited phonons. Note moreover that no switching is observed at the resonances corresponding to the TO phonons, even though the absorption of IR radiation is considerably stronger at the TO resonances, see Fig. 2a. This observation directly confirms the proposed switching mechanism is a result of resonant pumping of specific excitation - LO phonon modes - and not from mere thermal expansion of the lattice. Stroboscopic pump-probe experiments performed using a different laser source prove indeed that the mid-IR excitation can trigger magnetization dynamics on the time scale shorter than the usual precessional period (see Extended Data Fig. S6). Also in these experiments, the observed dynamics do not depend on the pump polarization, and shows the maximum amplitude of the excited precession at the wavelength of 14 μm, thus also proving that the excitation mechanism is the same as in switching experiments.

Figure 2b shows how the normalized switched area increases with an increase of pump fluence. Note that the threshold appearance of switched long-lived magnetic domains after irradiation is determined by a balance between the stable domain size and the diameter of the laser spot. On the other hand, further increase of the pump fluence above 2 J·cm$^{-2}$ leads to a local damage of the film[23]. Then to demonstrate the equivalency of switching starting from various initial magnetic phases, we repeat these experiment starting with different initial direction of magnetization, set by a brief application of the external magnetic field. Figure 3 shows the acquired images of the switched magnetic patterns for four initial magnetizations along the image diagonals.

To elucidate the role of the LO-phonon, we must note that the symmetry of the (001) garnet film differs from the one in the corresponding bulk crystal[24]. Due to the growth procedure, the film lacks the center of symmetry in the direction along the normal to the sample, while in the sample plane the original high symmetry survives. This leads to an efficient coupling of the LO mode to the TO ones[10], leading to a shift in the harmonic potential in the plane of the film, and thus creating an effective strain.

In order to understand the domain pattern induced by the pump pulse, we assume that the induced shift of the potential has, similar to the laser spot, a Gaussian profile. If $u$ is the displacement vector, the induced strain is given by

$$\varepsilon_{ij} = \frac{1}{2}\left(\frac{\partial u_i}{\partial x_j} + \frac{\partial u_j}{\partial x_i}\right), \tag{1}$$

where $x_i$ and $x_j$ are coordinates in the film plane. While the complete derivations can be found in the Methods, here we mention that the strains are assumed to be limited to the film plane (given by $x$–$y$ coordinates), so that the magnetoelastic energy[25] can be written as

$$E_{me} = b_1\left(\varepsilon_{xx} m_x^2 + \varepsilon_{yy} m_y^2\right) + 2b_2 \varepsilon_{xy} m_x m_y \tag{2}$$

In the limit of an isotropic sample, $b_1 = b_2$. Using the magnetoelastic energy ($E_{me}$) in the micromagnetic simulations (see Methods and Extended Data Fig. S7), we show that a single picosecond-long strain pulse leads to a selective reversal of four different areas in a homogeneously magnetized sample (see Fig. 3c). Interestingly, at the center of the spot, where the excitation has the highest amplitude, the reversal is absent. This can be understood from the first term in Eq. 2, which is isotropic and thus does not modify the state of the magnetic system. In contrast, the second term has quadrupolar symmetry and is directly responsible for the creation of the peculiar domain pattern. It is thus clear that our simple model fully reproduces the observations, and therefore confirms the switching mechanism.

In summary, here we demonstrate how to use a resonant excitation of LO phonon mode in a film of magnetic garnet in order to permanently switch its magnetization. Because of the growth-induced symmetry breaking strong anharmonic interaction between vibrational modes takes place.

Such interaction induces a displacement of lattice sites and therefore a strain pattern at the excitation area. As a result, we show that pumping of garnet films with mid-infrared pump pulses at either 14 μm or 22 μm wavelength switches the magnetic order of the film into a peculiar quadrupolar pattern. This pattern is rather counter-intuitive and unambiguously proves the switching mechanism. The switching is absent at the frequencies corresponding to TO phonons even though the absolute absorption is considerably stronger at these frequencies. This proves that the absorption as such does not play the role in the switching.

As an outlook, an ultrafast modification of the crystal field environment and thus of magnetocrystalline anisotropy, may become the most universal way to manipulate the magnetization. Magneto-elastic interactions are present in all materials and thus can be used everywhere, for example in antiferromagnets. On the other hand, the questions of magnetocrystalline anisotropy are extremely important because the anisotropy is the key point behind the stability of magnetization, for example in all magnetic memories.

**Acknowledgments** We acknowledge support from the grant of the Foundation for Polish Science POIR.04.04.00-00-413C/17-00. We gratefully acknowledge the Nederlandse Organisatie voor Wetenschappelijk Onderzoek (NWO-I) for their financial contribution, including the support of the FELIX Laboratory. We thank A. Maziewski for fruitful discussions as well as Th. Rasing for continuous support. We gratefully acknowledge Y.-L. Mathis for support at the IR1 beamline of the Karlsruhe Research Accelerator (KARA).

**Author Contributions** A.S. and A.K. conceived the project. A.S. performed the magneto-optical imaging together with C.S.D. D.A., K.S., A.C. performed time-resolved magnetization dynamics measurements. K.S.R, A.V.B. performed infrared ellipsometry measurements. C.S.D. performed the micromagnetic simulations. A.S., A.K., A.V.B. and A.V.K. jointly discussed the result and wrote the manuscript with contributions from all authors.

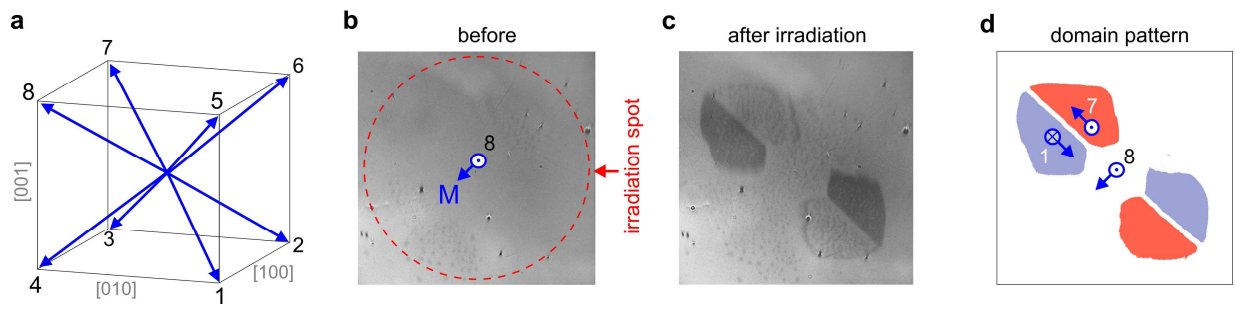

**Figure 1 | Ultrafast strain-induced switching of the magnetization in a cubic garnet.** (**a**) Four easy axes the magnetization in relation to the cubic crystal symmetry. (**b**) The image of initial state of magnetic domain structure with magnetization along 8 direction along [1-11] axis (large domain) and 4 direction along [11-1] axis (small domains) (**c**) The image of magnetic domain structure after radial strain generated by the pump pulse 0.5 ps and the fluence of 1.2 J·cm$^{-2}$. The photon energy was 87 meV (λ = 14 μm). (**d**) The visualization of domain pattern after image processing from (**c**). Switched blue and red area correspond to the orientation of the magnetization along the 1 and 7 axes. The image is 210×210 μm$^2$ large.

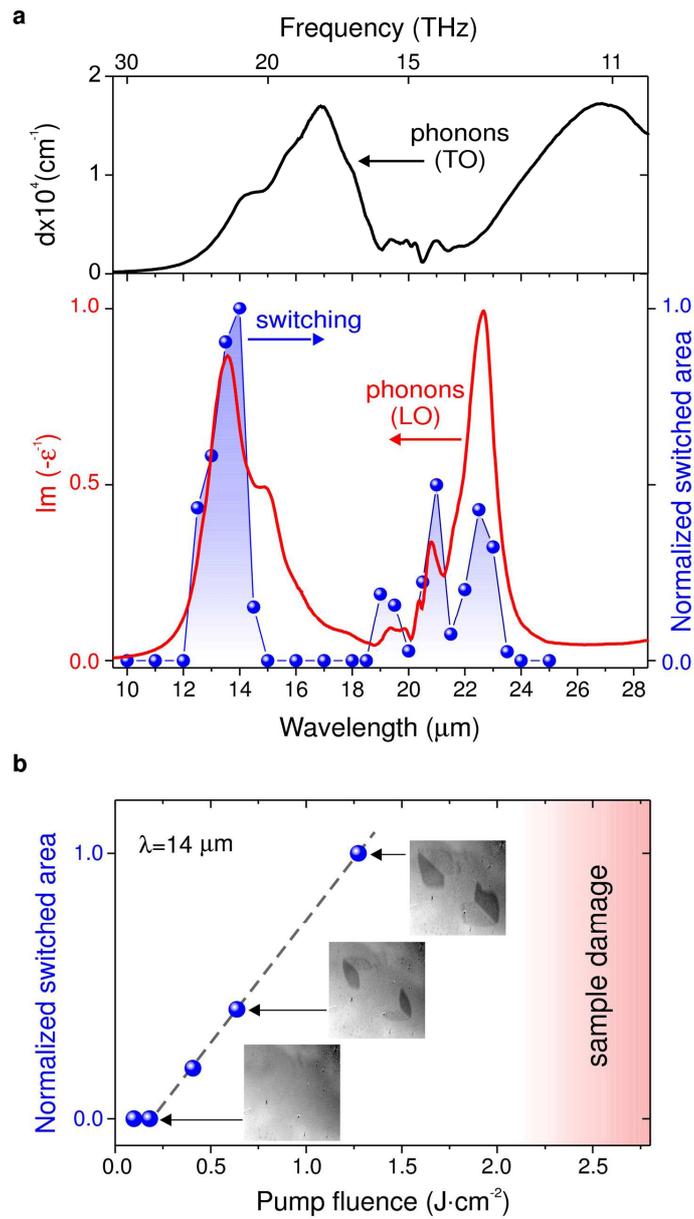

**Figure 2 | Resonant phonon modes excitation of magnetization switching. (a)** The graphs show the spectral dependence of the magnetization switching and optical phonon modes defined by loss function Im(-1/ε) and absorption $d$ in YIG film. The normalized switched area is calculated as the ratio of the switched area to the maximal switched area obtained for 14 μm laser irradiation with fluence 1.2 J·cm$^{-2}$ (laser spot was marked by red circle in Fig. 1b). **(b)** Fluence dependence of the normalized switched area for 14 μm pump laser irradiation. Dashed lines are guides to the eye. The inset show the images magnetic domains after switching with irradiation of different pump fluence patterns. The threshold of burning (local damage of the sample) in garnet was detected empirically.

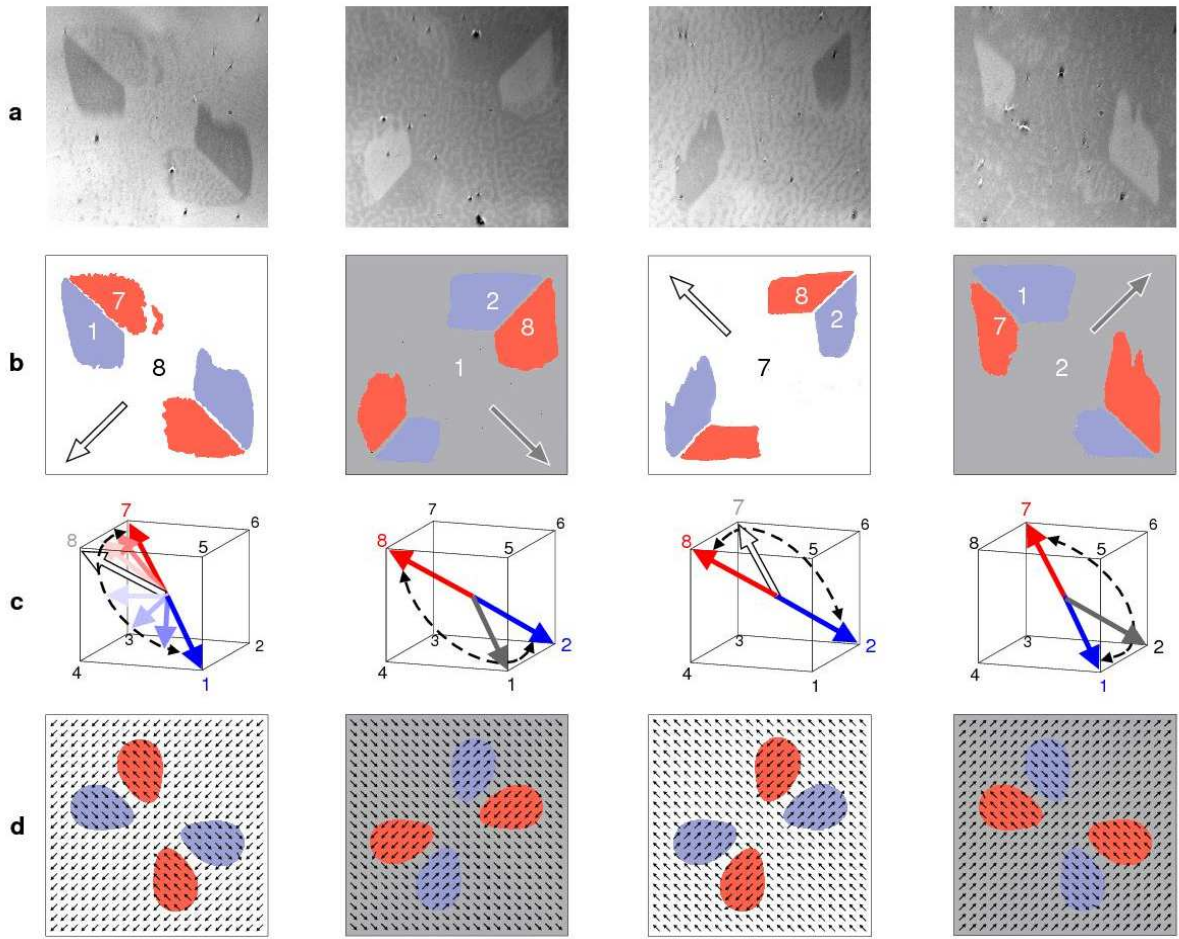

**Figure 3 | Multi-phase phononic switching in ferrimagnets.** Eight magnetic phases in relation to the cubic symmetry in garnet. (**a**) Images of strain-induced switching in domains for different four initial magnetization orientations (8, 1, 7 and 2). The fluence of the pump pulse was 1.2 J·cm$^{-2}$ for the wavelength of $\lambda$ = 14 μm. (**b**) Switched patterns of magnetic domains obtained from images in (a). (**c**) The panel shows schematically pathways of switching between different magnetic phases defined in Fig. 1a. (**d**) Simulations of switching with different initial orientation of the magnetization corresponding to the images in panel (a). The arrows and the red and blue area indicate the orientation of the magnetization after the application of the strain pulse. (**d**) Simulations of switching with different initial orientation of the magnetization corresponding to the images in panel. The arrows and the red and blue area indicate the orientation of the magnetization in the *x-y* plane after the application of the single strain pulse with duration 1 ps (see materials and methods).

**METHODS**

**Materials.** The measurements were performed on different YIG films with similar composition $(YCa)_3(FeCoGe)_5O_{12}$ and the thickness within 4–8 μm range. The garnets have been grown by liquid phase epitaxy on gadolinium gallium garnet (GGG) (001)-oriented substrates with both nominally flat surfaces and miscut at an angle of few degrees[26]. However, for better visualization of different magnetization states at domains, the images obtained on a sample grown on a 4° miscut GGG substrate with thickness 7.3 um are shown, though all results obtained are also equivalent to those obtained on the samples without miscut. The saturation magnetization at room temperature was about 7 G and the Néel temperature was 445 K. The Gilbert damping measured using ferromagnetic resonance technique was α=0.2. At room temperature the samples have both magnetocrystalline ($K_I \approx -10^4$ erg·cm$^{-3}$) and uniaxial ($K_U \approx -10^3$ erg·cm$^{-3}$) anisotropy, which were measured by means of both ferromagnetic resonance and torque magnetometry[27,28]. The easy axes of the magnetization are approximately along <111>-type directions. An additional feature of samples grown using liquid phase epitaxy is the growth-induced symmetry breaking along the film normal[24], that assures strong anharmonicity of the corresponding modes.

**Methods.**
*A. Magneto-optical imaging with FELIX pump pulses irradiation.* To investigate the magnetization switching in YIG films, we employed the technique of magneto-optical polarized microscopy using a pump laser pulse excitation at FELIX (Free Electron Lasers for Infrared eXperiments). The magnetic domain structure of the studied garnet films was visualized using magneto-optical polarized microscopy. The linearly-polarized light source was a LED lamp, with its output routed into the optical path of the probe beam. The LED light then transmitted through the YIG sample and was gathered with an objective before passing an analyzer and hitting the matrix of the CCD camera. The magnetic contrast in such a polarized microscope comes from the fact that magnetic domains with different out-of-plane magnetization components will give different Faraday rotation angle, and thus light passing through them will acquire different polarization, which can be easily detected on the CCD chip after filtering. We observed the final state the image of magnetic patterns, acquired no sooner than several milliseconds after laser excitation.

The magnetic domain structure was studied before[29,30,16-18]. Due to the miscut of the substrate we can directly observe four main magnetization states (1, 4, 5 and 8) as shown in Extended Data Fig. S1. Such a configuration shows the breaking of the degeneracy between "5" and "8" or "1" and "4" domains. In this case, "8" and "1" - domains are always larger than the small "5" and "4" domains. This can be shown by application of an external magnetic field in different in-plane directions <110>–type, with corresponding projections of magnetization on the sample plane. The initial "8" state was achieved by an external magnetic field of 10 mT applied along the [1-10] direction for a few seconds. When the field is removed, the sample turns into a state with "8" (large domains) and "4" (small domains) domains. The same scenario we also used for other initial states (1, 7 and 2) when the magnetic field was applied along the [110], [-110] and [-1-10] directions (see Fig. 3). Without any magnetic field, the magnetic pattern stays unchanged for at least several days due to the non-zero coercivity in garnets. All switching experiments were done in zero applied magnetic field and at room temperature.

The central wavelength of the pump pulses, with duration of the range 0.5–1 ps (depending on wavelength), was varied in the spectral range of 10–30 μm with their bandwidth experimentally tunable in the range of 0.5–2.0% (typically <1% bandwidth used in our experiments). The laser beam was focused to a spot of diameter of about 200 μm onto the surface of the garnet film. The images of magnetic domains were taken before (reference) and after the pump excitation. Taking the difference between the images underlines the laser-induced changes - therefore, such difference images are better suitable for a detailed analysis (see Extended Data Fig. S2).

*B. Spectroscopic ellipsometry.* We have used synchrotron-based spectroscopic ellipsometry to accurately determine the infrared phonon spectra of YIG on GGG. The ellipsometric measurements in the frequency range 7.5 meV to 1 eV (60 to 8000 cm$^{-1}$) were performed at room temperature using homebuilt ellipsometers in combination with Bruker IFS 66v/S and Vertex 80v Fourier transform infrared spectrometers. The measurements in the far infrared (FIR, 7.5 to 88 meV, 60 to 700 cm$^{-1}$) utilized synchrotron edge radiation of the 2.5 GeV electron storage ring at the IR1 beamline of the Karlsruhe Research Accelerator (KARA) at the Karlsruhe Institute of Technology, Germany. The ellipsometric parameters $\Psi$ and $\Delta$ measured at several angles of incidence ranging from 70º to 80º define the complex ratio $r_p/r_s = tan(\Psi) \, exp(i\Delta)$, where $r_p$ and $r_s$ are the complex Fresnel coefficients for light polarized parallel and perpendicular to the plane of incidence, respectively[22]. The real and imaginary parts of the complex dielectric function $\varepsilon(\omega) = \varepsilon_1(\omega) + i\varepsilon_2(\omega)$ in Fig.S3 are derived by a direct inversion of $\Psi$ and $\Delta$ assuming semi-infinite bulk isotropic behavior of the YIG film. The transverse optical (TO) phonons appear as peaks in $\varepsilon_2(\omega)$, while the longitudinal optical (LO) phonons cause peaks in the imaginary part of the negative inverse of the dielectric function, i.e. loss function $Im(-1/\varepsilon)$.

The influence of reflection at the film-substrate interface on the obtained phonon spectra is negligible and can only be noticed in the range of around 500 cm$^{-1}$ (indicated by dotted lines). The film thickness was determined by periodic oscillations in the transparency window of the ellipsometric spectra due to interference of multiple reflections within the film (0.2 to 0.5 eV, not shown). The optical constants of GGG were obtained from independent spectroscopic ellipsometry measurements and also used to process $\Psi(\omega)$ and $\Delta(\omega)$ spectra measured on the YIG film on the substrate at different incident angles. In order to confirm that the solid lines in Extended Data Fig. S3 correspond to the true dielectric function for isotropic YIG, the data was also analyzed through a best-match single-film model calculation procedure, as implemented in the Woollam WVASE32® data acquisition and analysis software[31]. With the knowledge of the optical constants of GGG and the YIG layer thickness fixed at 7.3 µm, we made sure of rapid convergence of the wavelength-by-wavelength regression procedure to the experimental ellipsometric spectra with the same dielectric function of YIG as shown in Extended Data Fig.S3.

Using the dielectric function of YIG, the absorption coefficient spectrum $d(\omega)$ of Extended Data Fig. S4 (black curve) was obtained. The result is consistent with the absorption spectrum of bithmuth/gallium-substituted YIG[6]. By means of ellipsometry we were able to measure not only the absorption spectrum of infrared-active TO phonons, but also the loss function spectrum (red curve), which reveals strong peaks at 445 cm$^{-1}$ and 730 cm$^{-1}$ attributed to LO-phonon modes in YIG.

*C. Time-resolved pump-probe magnetization dynamics measurements using NIR laser pulses*
The intense pump pulses with photon energy in the mid-infrared spectral range used in this experiment were generated using difference frequency generation in a GaSe crystal (0.35 mm), using the output beams of two commercially available, independently tunable optical parametric amplifiers (OPAs) integrated into a single housing (Light Conversion, TOPAS-Twins). The OPAs were pumped by a commercially available amplified Ti:Sapphire laser system (Coherent, Astrella) delivering pulses at 1 kHz repetition rate with a duration of 100 fs with a photon energy of 1.5 eV. The OPAs were seeded by the same white light generated in a sapphire crystal, which ensures separately tunable, but phase-locked output pulses with photon energies in the range of 0.45 eV to 1 eV. As a result, when these output pulses are mixed in the GaSe crystal, the generated mid-IR pulses are carrier envelope phase stable and their energy lies in the range from 65 meV to 250 meV with an average pulse duration of around 200 fs. In the experiments, the mid-infrared pulses were focused onto the sample surface to a spot with a diameter of about 150 µm, using an off-axis parabolic mirror. The ensuing coherent spin precession is measured, in a conventional pump-probe scheme, by tracking the polarization rotation, imprinted by the magneto-optical Faraday effect, on co-propagating probe pulses at the photon energy of 1.5 eV.

*D. Micromagnetic simulations.* Micromagnetic modelling was performed using the finite-difference time-domain software Object-Oriented Micromagnetic Framework (OOMMF) release 2.0a1[32]. Our sample volume is of size 500 ×500 ×7.5 μm$^3$, and discretized with cells of size 2 ×2×7.5 μm$^3$. We impose uniform magnetization of saturation equivalent to $4\pi M_S \sim 100$ G. Throughout all calculations, the Landau-Lifshitz-Gilbert equation was solved using the Runge-Kutta-Fehlberg integration method.

The effective magnetic field consists of several terms. We neglect the demagnetization field since the magnetization is relatively small. We also neglect the exchange field because the cell-size is three orders of magnitude larger than the exchange length. We note this approach has been successfully used before macroscopic length scales e.g. while calculating magnetostatic spin-wave dispersion relations[33]. We use a cubic magneto-crystalline anisotropy field of strength $K_1 = -1$ kJ/m$^3$ (equivalent to $10^4$ erg/cm$^3$) with easy axes along the [110], [1-10] and [001] directions. We also insert a uniaxial magneto-crystalline anisotropy field of strength $K_u = -500$ kJ/m$^3$ with its hard axis along the [001] direction. This substantially-stronger uniaxial anisotropy field was inserted to compel the magnetization to stay close to the sample plane, thus allowing us to consider the problem in two dimensions only. As a result, the sample effectively has two easy axes along the [110] and [1-10] directions. The magnetoelastic field was incorporated using the extension developed by Yuhagi *et al.*[34], and will be discussed later.

**Pump-induced demagnetization.** The magnetic pattern deduced from the images after irradiation for wavelength 17 μm, corresponding the TO-phonons peak, does not show the switching of the large magnetic domains. At the maximum intensity in the center of the Gaussian profile, a multidomain pattern is observed, which is a fingerprint of laser-induced demagnetization by heating at the center of the image (see Extended Data Fig. S5). It thus proves that the switching resulting in the peculiar spatial pattern indeed occurs via a nonthermal mechanism. We should also note that the magnetization switching does not depend on the pump polarization within the whole spectral range of the measurements.

**Ultrafast excitation of magnetic precession.** To prove that the induced strains can indeed trigger the magnetization dynamics on the time scale shorter than the usual precessional period, we carried out stroboscopic experiments on a different laser source. While of insufficient pulse energy to excite the switching, the results clearly show an almost immediate onset of magnetic precession started by the mid-IR laser pulses (see Extended Data Fig. S6). Also in these experiments, the observed dynamics does not depend on the pump polarization within the whole spectral range of the measurements. Changing the wavelength of the excitation demonstrates the maximum amplitude of the excited precession at the wavelength of 14 μm, thus also proving that the excitation mechanism is the same as in switching experiments.

**Theoretical model of strain-induced switching of magnetization.** We assume that the anharmonic interaction of the phonon modes results in the shift of the in-plane coordinates proportional to the local intensity of the excitation. Such modification of the local coordinates is proportional to the laser intensity and thus follows the Gaussian profile of the laser spot. Though microscopically of different origin, macroscopic distribution of the strains will be equivalent to the case of a non-uniformly heated object (e.g. cylinder) with axially-symmetric temperature distribution $T(r)$, that was studied in the past[35].

The displacement in this case is given by

$$u = q\frac{1+\sigma}{1-\sigma}\left[\frac{1}{r}\int_0^r T(r)r\mathrm{d}r + (1-2\sigma)\frac{r}{R^2}\int_0^R T(r)r\mathrm{d}r\right],$$

where $q$ is the thermal expansion coefficient, $\sigma$ is the Poisson ratio and $R$ is the radius of the cylinder. By assuming that the effect of the absorbed laser fluence can be treated as a spatially-Gaussian "temperature" i.e. $T(r) = Ae^{-\frac{r^2}{2a^2}}$, we obtain the final form of the displacement

$$u = \frac{Aqa^2}{3}\frac{1+\sigma}{1-\sigma}\left[\frac{1}{r}\left(1 - e^{-\frac{r^2}{2a^2}}\right) + (1-2\sigma)\frac{r}{R^2}e^{-\frac{R^2}{2a^2}}\right],$$

where $r = \sqrt{(x-x_0)^2 + (y-y_0)^2}$, $x_0 = y_0 = 250$ μm and $a = 50$ μm is the Gaussian standard deviation of the pulse. Therefore the full-width half-maximum of the pulse is 118 μm. All other strain elements are set to zero. With $R$ tending to infinity (i.e. we consider not a cylinder but rather a plane), the second term goes to zero. The constant term before the brackets we denote as β:

$$u = \beta\frac{1}{r}\left(1 - e^{-\frac{r^2}{2a^2}}\right)$$

Going to Cartesian $(x,y)$ coordinates and calculating the derivatives gives the spatial distribution of strain

$$\varepsilon_{xx}(x,y) \equiv \frac{\partial u_x}{\partial x} = \frac{\beta}{x^2+y^2}\left[\left(1 - e^{-\frac{x^2+y^2}{2a^2}}\right)\left(1 - \frac{2x^2}{x^2+y^2}\right) + \frac{x^2}{a^2}e^{-\frac{x^2+y^2}{2a^2}}\right],$$

$$\varepsilon_{yy}(x,y) \equiv \frac{\partial u_y}{\partial y} = \frac{\beta}{x^2+y^2}\left[\left(1 - e^{-\frac{x^2+y^2}{2a^2}}\right)\left(1 - \frac{2y^2}{x^2+y^2}\right) + \frac{y^2}{a^2}e^{-\frac{x^2+y^2}{2a^2}}\right],$$

$$\varepsilon_{xy}(x,y) \equiv \frac{1}{2}\left(\frac{\partial u_x}{\partial y} + \frac{\partial u_y}{\partial x}\right) = \frac{-\beta xy}{x^2+y^2}\left[\frac{2}{x^2+y^2}\left(1 - e^{-\frac{x^2+y^2}{2a^2}}\right) - \frac{1}{a^2}e^{-\frac{x^2+y^2}{2a^2}}\right],$$

The magnetoelastic energy is then written as

$$E_{me} = b_1(\varepsilon_{xx}m_x^2 + \varepsilon_{yy}m_y^2) + 2b_2\varepsilon_{xy}m_xm_y,$$

The spatial distribution of the first (symmetric) and second (antisymmetric) terms is shown in Extended Data Fig. S7. To obtain the ground magnetic state of the sample, we uniformly tilted the magnetization slightly from a selected easy axis. The Gilbert damping parameter $\alpha$ was set to 0.5, and the magnetization was allowed to relax to its equilibrium state (defined as complete when the maximum d$M$/dt across all cells became smaller than 0.01 °/ns). Dynamic simulations were performed using the final magnetic state obtained from the prior relaxation process. Calculations were performed with $\alpha = 0.2$, and the magnetization was saved every 100 fs for a total duration of 50 ps. The magnetoelastic field pulse $h_{me}(r,t) = h_{me,0} \cdot h_{me}(r) \cdot h_{me}(t)$ is of amplitude $h_{me,0} = 8.2$ T, and is split in to time- and space-varying functions. In time, the pulse has Gaussian character

$$h_{me}(r) = \exp\left(-\frac{[t-t_0]^2}{[\tau^2/(4ln2)]}\right),$$

where $t_0 = 2$ ps is the pulse's delay in time, and $\tau = 1$ ps is the pulse's full-duration half-maximum. In space, the pulse is of the form $h_{me}(r) = -\frac{\delta U_{me}}{\delta M}$, where $U_{me}$ is the magnetoelastic energy density

$$U_{me}(r) = \frac{b_1}{M_s^2}\sum_i M_i^2\varepsilon_{ii} + \frac{b_2}{M_s^2}\sum_i\sum_{j\neq i} M_iM_j\varepsilon_{ij}.$$

The indices $i,j = 1,2,3$ correspond to the Cartesian coordinates $x$, $y$ and $z$ respectively. We assume that the sample is isotropic, and so $b_1 = b_2 = -1$. The magnetoelastic strain terms $\varepsilon_{i,j}$ are obtained by considering the sample to be non-uniformly heated in the $x$–$y$ plane by a Gaussian pulse.

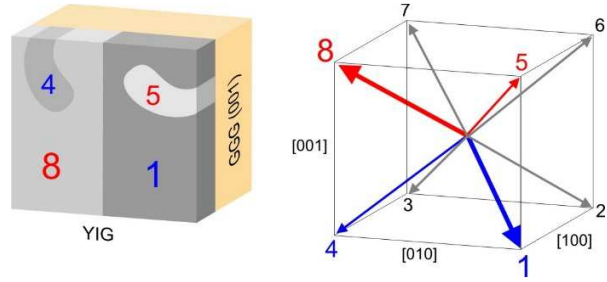

**Extended Data Fig. S1** | Magnetic domains pattern with four magnetization states in relation to the crystal symmetry in YIG/GGG(001). The large stripe-like domains are formed that have magnetizations along the [1-11] and the [11-1] axes. Within them, small domains are found that possess magnetization along the [111] and the [1-1-1] axes. Magnetization orientations in the domains and type of the domain structure have been identified following the procedure explained in[29,16].

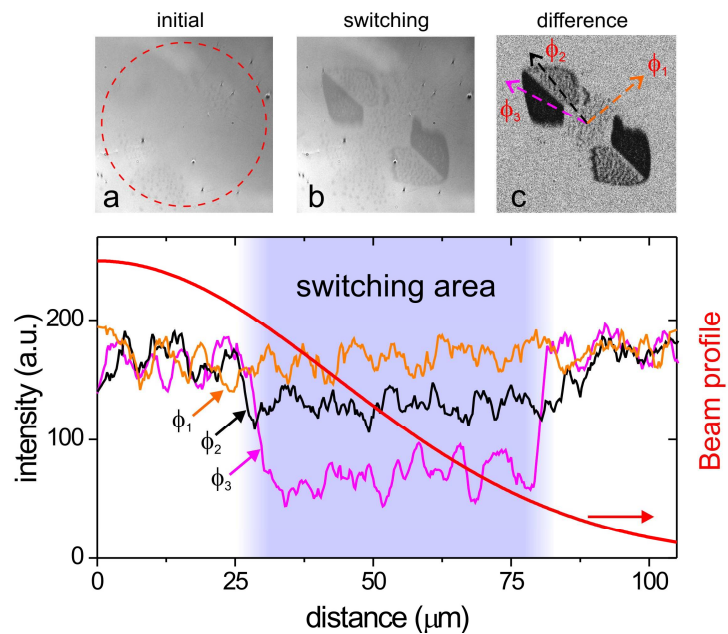

**Extended Data Fig. S2** | Top panel from left to right shows the images of domain structures before the laser excitation (a), after irradiation at pump wavelength 14 μm (b) and difference these images (c). The laser spot was marked by red dashed circle. The initial domain structure was observed by applying an external in-plane magnetic field 10 mT along the [1-10] axis for one second. The graph shows the spatial profiles of the intensity from difference image with magnetic contrast after irradiation. The red solid line represents the beam profile (half part) of irradiation pulse The images are 210×210 μm$^2$ large.

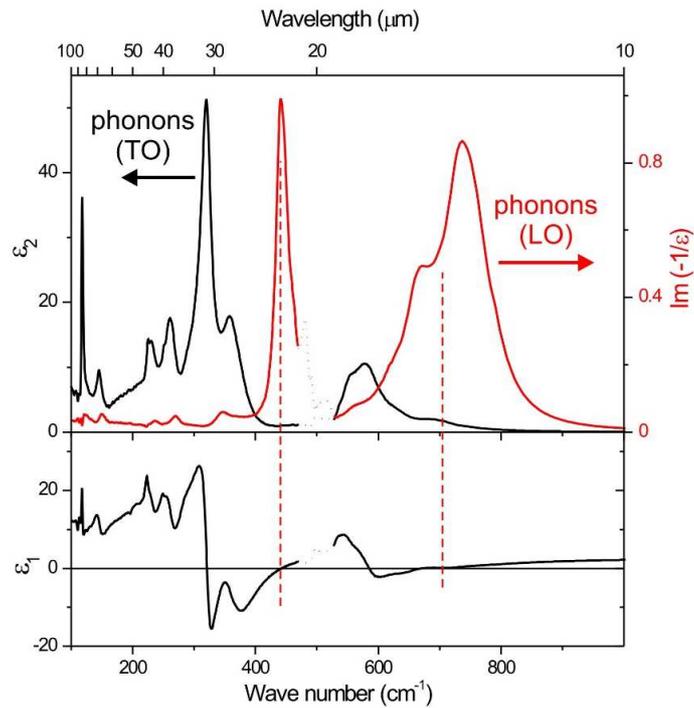

**Extended Data Fig. S3 | Phonon spectra of YIG.** Complex dielectric function $\varepsilon(\omega) = \varepsilon_1(\omega) + i\varepsilon_2(\omega)$ (black) and the imaginary part of the dielectric loss functions $\mathrm{Im}(-1/\varepsilon)$ (red) of YIG obtained by direct inversion of ellipsometric spectra. Vertical dashed lines indicate zero crossing of $\varepsilon_1(\omega)$.

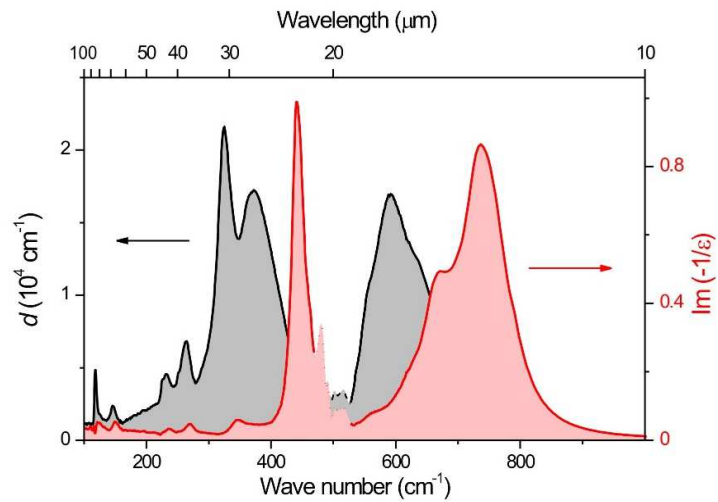

**Extended Data Fig. S4 | Infrared absorption and loss function.** Absorption coefficient (black) and imaginary part of the loss function (red) of YIG determined from ellipsometric spectra.

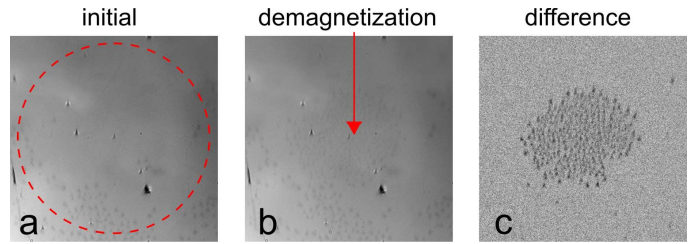

**Extended Data Fig. S5 |** The images of magnetic domain structures before the laser excitation (a), after irradiation at pump wavelength 17 μm (b) and difference these images (c). The laser spot was marked by red dashed circle.

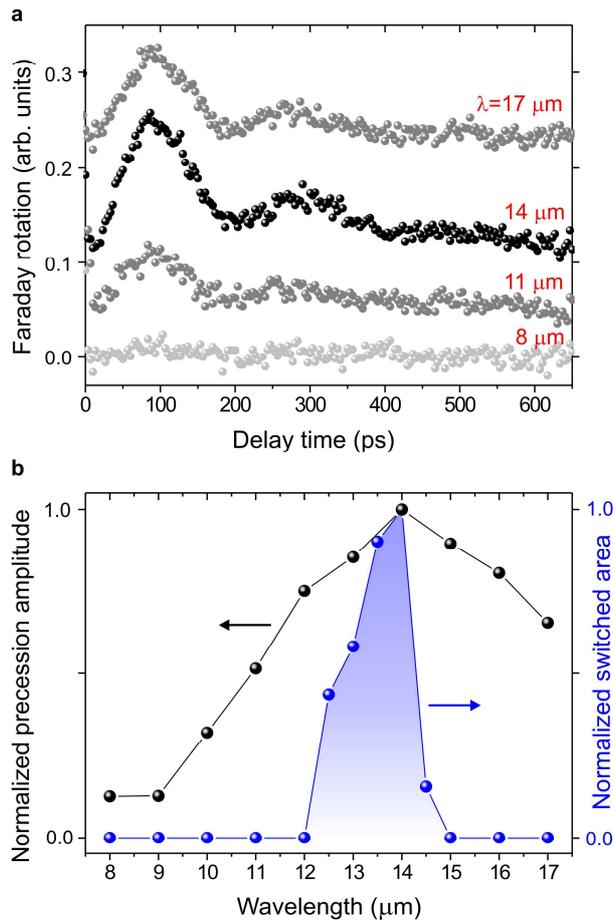

**Extended Data Fig. S6 |** Time-resolved magnetization precession induced by the femtosecond pump pulses in YIG film. The out-of-plane component of the magnetization is detected with the help of time-resolved magneto-optical Faraday rotation. (a) The magnetization precession measured using different pump wavelengths in the range from 8 μm to 17 μm. (b) Dependence of the normalized precession amplitude on the pump wavelength (black points) and magnetization switching (blue points) from Fig. 2. The normalized precession amplitude is calculated as the ratio of the precession amplitude to the maximal amplitude obtained for 14 um pump laser-induced with pump fluence 0.06 J·cm$^{-2}$.

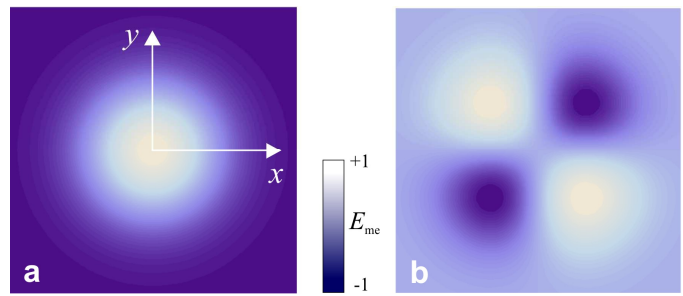

**Extended Data Fig. S7 |** Magneto-elastic energy distribution created by the excitation of the phonon mode: (a) symmetric part with $b_2$=0 (the arrows on the axes indicate $x$–$y$ coordinates), and (b) antisymmetric part with $b_1$=0.